\documentclass{PoS}
\usepackage{amsmath}

\title{Multi-component dark matter from a hidden gauged SU(3)}

\ShortTitle{Multi-component dark matter from a hidden gauged SU(3)}

\author{\speaker{Stephen Godfrey} and Alexandre Poulin 
\\
        Ottawa-Carleton Institute for Physics, Carleton University, 1125 Colonel By Drive, 
Ottawa, Ontario K1S 5B6, Canada\\
        E-mail: \email{godfrey@physics.carleton.ca}, \email{alexandre.poulin@carleton.ca}}

\abstract{We studied Dark Matter (DM) phenomenology with multiple DM species consisting of both scalar 
and vector DM particles in the Hidden Gauged SU(3) model of Arcadi et al. 
Because of the large parameter space in the Hidden Gauged SU(3) 
model we restrict ourselves to three representative benchmark points, 
each with multiple DM species. The relic densities for the benchmark points were found using a program 
developed to solve the coupled Boltzmann equations for an arbitrary number of interacting DM species 
with two particles in the final state. For each case, we varied the mass of the DM particles and then 
found the value of the dark SU(3) gauge coupling that gave the correct relic density. We found that 
in some regions of parameter space, DM would be difficult to observe in direct detection experiments while it 
would be easier to observe in indirect detection experiments while for other regions of parameter 
space the situation was reversed.  Thus, measurements from both types of experiments complement 
each other and could help pinpoint the details of the hidden SU(3) model.}

\FullConference{XXIX International Symposium on Lepton Photon Interactions at High Energies - LeptonPhoton2019\\
		August 5-10, 2019\\
		Toronto, Canada}

\begin{document}

\section{Introduction}

There is evidence that dark matter (DM) makes up a significant fraction of the total matter
in the universe but virtually nothing is known about it except its relic abundance. While  weakly interacting 
massive particle (WIMP) DM is an attractive paradigm, the simplest models of WIMP DM 
are either significantly constrained or entirely ruled out \cite{Arcadi:2017kky}.  
In this contribution, we describe the phenomenological
implications of a more complicated model, the multispecies 
hidden SU(3) model of Arcadi {\it et al} \cite{Arcadi:2016kmk}. A more detailed report of this
analysis is given in \cite{Poulin:2018kap}. 
 For the parameter space
we focus on, the model has 4-7 DM species consisting of scalar and vector DM.  
We find that   the vector particles can more readily be
observed in direct detection while  the scalar particles are more likely to be 
observed in indirect detection measurements.

\section{The Hidden Gauged SU(3) Model}

We study the phenomenology of the Hidden Gauged SU(3) model
 of Arcadi {\it et al.} \cite{Arcadi:2016kmk} which consists of spin-1 and spin-0 states.  
In their paper, they examined two representative limiting cases that made their numerical 
analysis more tractable.  We extend their studies to consider different points in the parameter 
space that leads to more complex DM scenarios \cite{Poulin:2018kap}.  

The  model consists of a gauged SU(3) which is fully broken by two complex scalar 
triplets, $\Phi_1$ and $\Phi_2$, so that all the new gauge bosons acquire a mass. 
These new scalars are not charged under the SM gauge groups and can only interact with the SM
through the SM Higgs doublet. To simplify the Lagrangian and insure 
additional stable states, a $\mathbb{Z}_2$ symmetry is also imposed such that
$\Phi_1\rightarrow-\Phi_1$ and $\Phi_2\rightarrow \Phi_2$, 
which has the effect of only including even powers of the scalar triplets in the Lagrangian.
Following Ref.~\cite{Arcadi:2016kmk}, 
the Lagrangian is given by:
\begin{equation}
\mathcal{L}=\mathcal{L}_{\rm SM}+\mathcal{L}_{\rm portal}+\mathcal{L}_{\rm hidden},
\end{equation}
where $\mathcal{L}_{\rm portal}$ and $\mathcal{L}_{\rm hidden}$ are the new DM pieces given by
\begin{align}
-\mathcal{L}_{\rm portal} & = V_{\rm portal}=\lambda_{H11}|H|^2|\Phi_1|^2+\lambda_{H22}|H|^2|\Phi_2|^2, \\
-\mathcal{L}_{\rm hidden} &=-\frac{1}{4}G_{\mu\nu}^a G^{\mu\nu a}+|D_\mu \Phi_1|^2+|D_\mu \Phi_2|^2 
- \frac{\lambda_2}{2}|\Phi_2|^4 \nonumber \\
& \qquad  -\lambda_3  |\Phi_1|^2|\Phi_2|^2
-\lambda_4|\Phi_1^\dag\Phi_2|^2 -\frac{\lambda_5}{2}\left[(\Phi_1^\dag\Phi_2)^2+h.c\right].
\end{align}
Here, $G_{\mu\nu}^a=\partial_\mu A_\nu^a -\partial_\nu A_\mu^a +\tilde{g}f^{abc}A^b_\mu A^c_\nu$ is the 
field strength tensor for the hidden SU(3) where $\tilde{g}$ is the gauge coupling and $f^{abc}$ are 
the SU(3) structure constants. 

In the scalar sector, working 
in the unitary gauge and because of the specific choices we made \cite{Poulin:2018kap},
the scalar mass matrix is a $5\times 5$ matrix where $\Phi^T=(h,  \varphi_1, \varphi_2, \varphi_3, 
\varphi_4 )$. Due to our choices, $\varphi_3$ and $\varphi_4$ cannot mix with any of the 
other scalars and because we do not have $CP$ violation in the scalar sector, they also cannot
mix with each other so both are mass eigenstates which we relabel as $\varphi_3=\mathcal{H}$ 
and $\varphi_4=\chi$.  The remaining $3\times 3$ mass matrix can be diagonalized using
three mixing angles.  For $\theta_2$ small, one can
define an effective mixing angle $\theta = \theta_1 + \theta_3$ which mixes $h$ and $\phi_2$.  
$\theta$ is constrained to be small by the Higgs boson properties \cite{Arcadi:2016kmk,Falkowski:2015iwa}.

In the vector sector, only $A^3$ and $A^8$ mix 
and 
the vector boson masses are given by:
\begin{equation}
m_{A^{1'},A^{2'}}=\frac{\tilde{g}}{2}v_2,\
m_{A^{4'},A^{5'}}=\frac{\tilde{g}}{2}v_1,\
m_{A^{6'},A^{7'}}=\frac{\tilde{g}}{2}\sqrt{v_1^2+v_2^2}, \
m_{A^{3'},A^{8'}}^2  =\frac{\tilde{g}^2}{6}\left(v_1^2+v_2^2\mp\sqrt{v_1^4-v_1^2v_2^2+v_2^4}\right).
\end{equation}
Where $v_1$ is the vev for the 3rd component of $\Phi_1$ and $v_2$ and $v_3$ are the
vevs for the 2nd and 3rd components of $\Phi_2$.
We can see that $A^{3'}_\mu$ is the lightest vector boson 
while $A^{8'}_\mu$ is the heaviest, or one of the heaviest. When $v_1=v_2$,
$A^{8'}_\mu$ becomes degenerate with $A^{6'}_\mu$ and $A^{7'}_\mu$.

The model has 3 global $\mathbb{Z}_2$ symmetries.  Due to these symmetries and the decay kinematics
resulting from the masses, the model will have a minimum of 4 stable DM species and a maximum
of 7.

Before applying the experimental constraints, we apply the usual theoretical constraints on
the parameters of the model: partial wave unitarity, the scalar potential being bounded from below, and
that we avoid alternative minima.  These details are outlined in Ref.~\cite{Poulin:2018kap}.

\section{Results and Discussion for the Hidden Gauged SU(3) Model}

The core calculation of our analysis is calculating the relic abundance when many species are present.
We adapt the coupled Bolzmann equations given by Dienes, Huang and Thomas \cite{DHT2018},
which can include an arbitrary
number of DM species, to include only $2\to 2$ and $1\to 2$ interactions.  
 Integrating the coupled
Boltzmann equations to obtain the relic density for multiple DM species is computationally 
intensive and because the parameter space of the Hidden Gauged SU(3) model is large, 
it is impractical to perform a complete scan of the parameter space.
Instead we explore the implications of three 
representative benchmark points. 
The parameter values  for  benchmark point A
were chosen to explore the region where one of the DM
 species could annihilate to SM particles through a Higgs resonance. Benchmark point B 
was chosen to explore a region where no resonant effects would occur. Finally, benchmark point 
C was chosen to explore the region where $v_1$ and $v_2$ have similar values and where there
are many more stable DM particles.

We vary the mass of one of the DM species and find the gauge coupling $\tilde{g}$ which 
results in the correct relic density.  We  vary the mass of $\chi$, $\mathcal{H}$, and $A^{1'}$ for 
benchmark point A, B, and C, respectively. 
Some of the hidden sector particles are stable because symmetry respecting decays
 are kinematically forbidden.
Thus, benchmark points A and B have 
4 stable DM species, while benchmark point 
C has 7 DM species.  The benchmark points
we consider are defined in Table~\ref{table:benchmark}.  We
ignore any unstable dark sector species by assuming that they would have decayed out long
before freezeout occurred.  Note that for benchmark point A, there is a disallowed region around 
half the Higgs mass because the gauge coupling required to give the correct relic abundance
violates the unitarity limit for gauge boson scattering.

\begin{table*}[t]
\begin{center}
\begin{tabular}{ |c|c|c|c| }
\hline
Parameters &  Scenarios A & Scenarios B & Scenarios C \\
\hline\hline
$m_{{A^1}'}$ & $300$ GeV & $300$ GeV & $200-500$ GeV  \\\hline
$m_{h_2}$ & $2500$ GeV & $2500$ GeV & $600$ GeV \\\hline
$m_{h_3}$ & $650$ GeV & $650$ GeV & $225$ GeV \\\hline
$m_{\mathcal{H}}$ & $1000$ GeV & $400 - 600$ GeV & $250$ GeV\\\hline
$m_{\chi}$ & $50 - 200$ GeV & $1000$ GeV & $250$ GeV \\\hline
$v_1/v_2$ & 10 & 10 & 1.2  \\\hline
DM & $\chi$, ${A^1}'$, ${A^2}'$, ${A^3}'$& $\mathcal{H}$, ${A^1}'$, ${A^2}'$, ${A^3}'$ & $\mathcal{H}$, $\chi$, ${A^1}'$, ${A^2}'$,\\
 &  & &  ${A^3}'$, ${A^4}'$, ${A^5}'$\\\hline
\end{tabular}
\end{center}
\caption{Input parameters for the benchmark points. $m_\chi$, $m_{\mathcal{H}}$, and $m_{{A^1}'}$ were allowed to 
vary for benchmark
point A, B, and C, respectively, along with the gauge coupling which was varied to obtain the correct relic density.
In all cases, we take the mixing in the scalar sector to only
be between $h_1$ and $h_3$ with a mixing angle of $\sin\theta=0.1$. We also set $m_{h_1}$ 
equal to the SM Higgs mass.}
\label{table:benchmark}
\end{table*}

\begin{figure}[t]
\begin{center}
$
\includegraphics[scale=0.15,keepaspectratio=true]{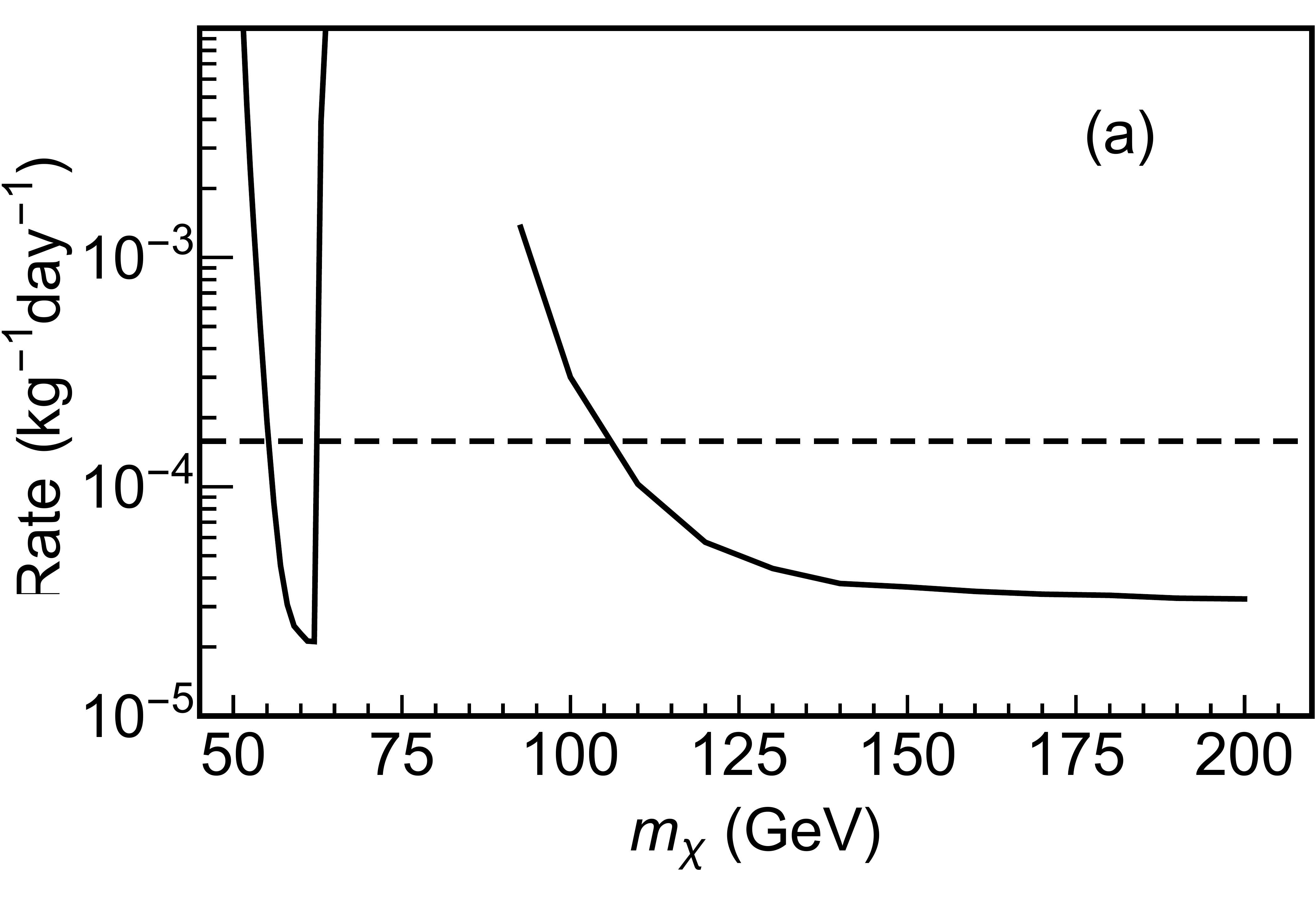} 
$
$\quad$
$
\includegraphics[scale=0.15,keepaspectratio=true]{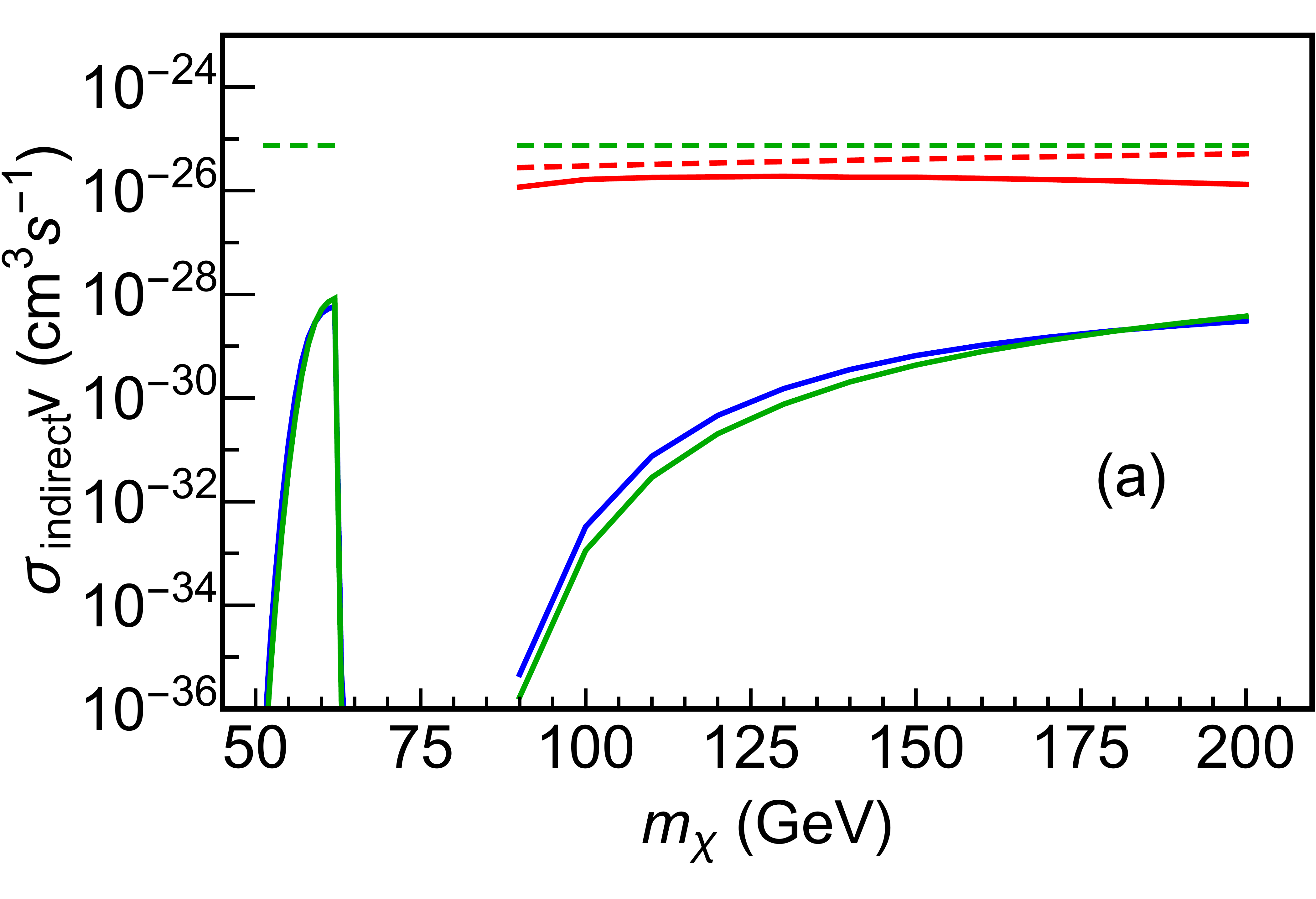} 
$
\end{center}
\begin{center}
$
\quad
\includegraphics[scale=0.145,keepaspectratio=true]{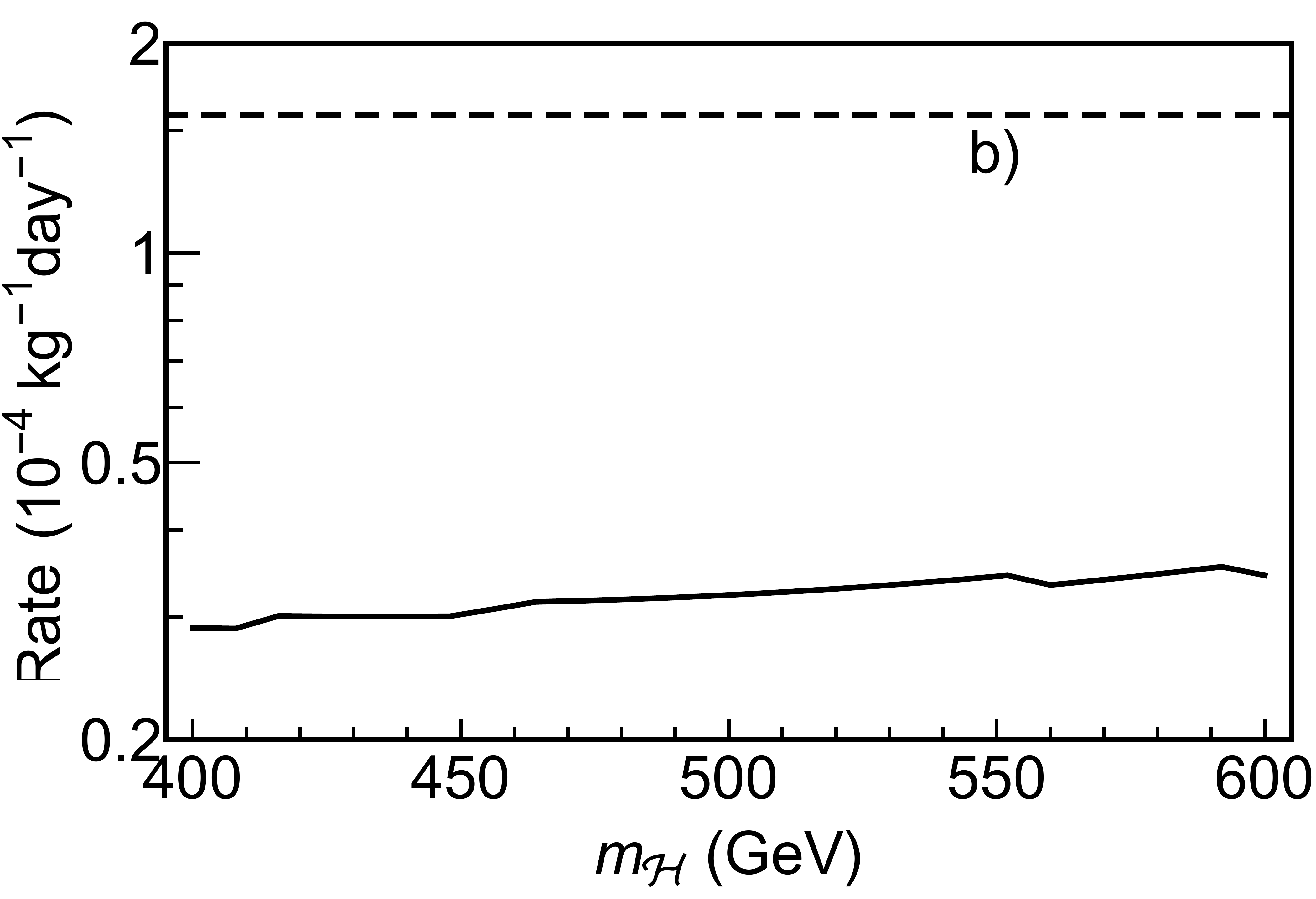}
$ 
$\quad$
$
\includegraphics[scale=0.155,keepaspectratio=true]{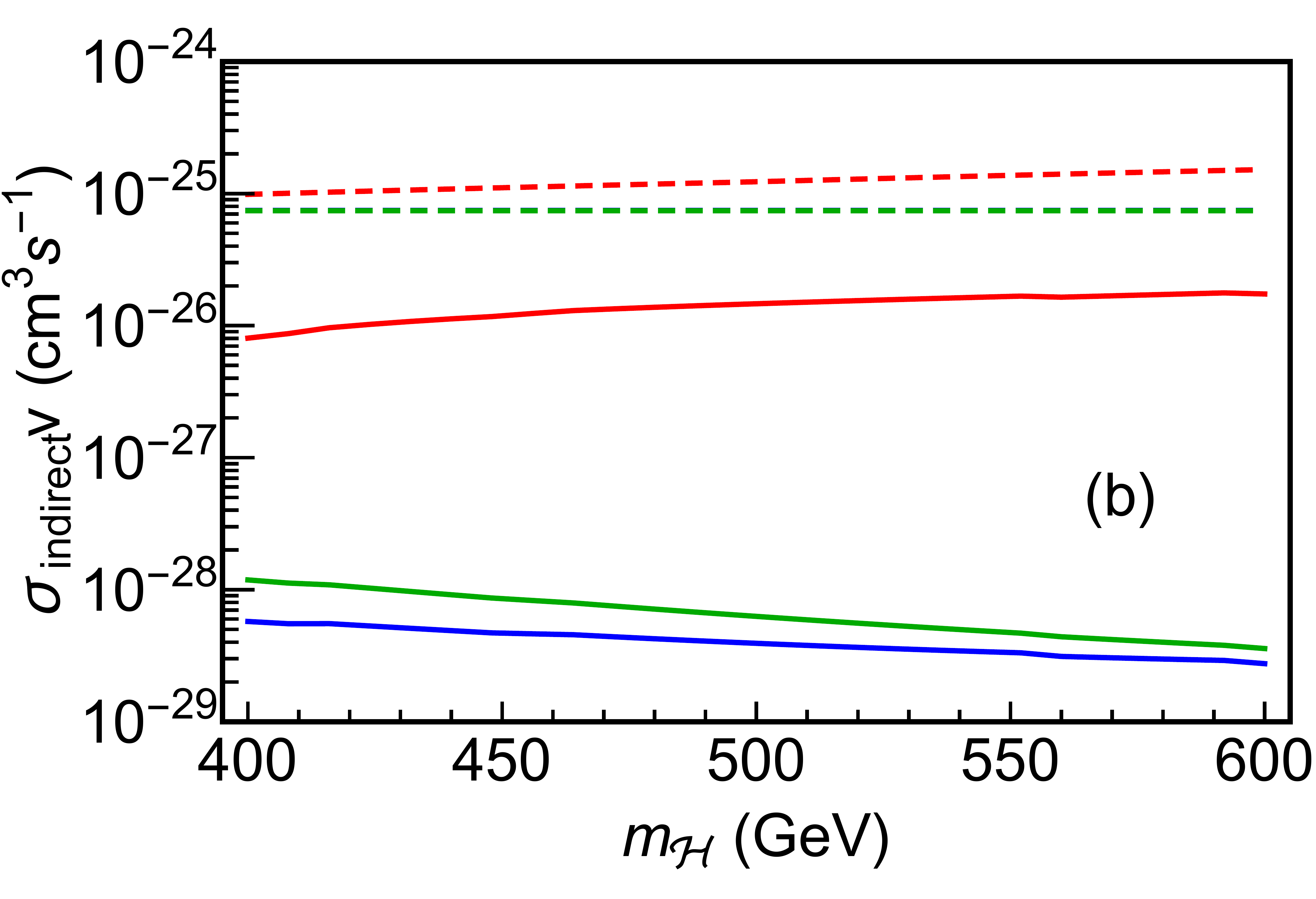}
$ 
\end{center}
\begin{center}
$
\includegraphics[scale=0.17,keepaspectratio=true]{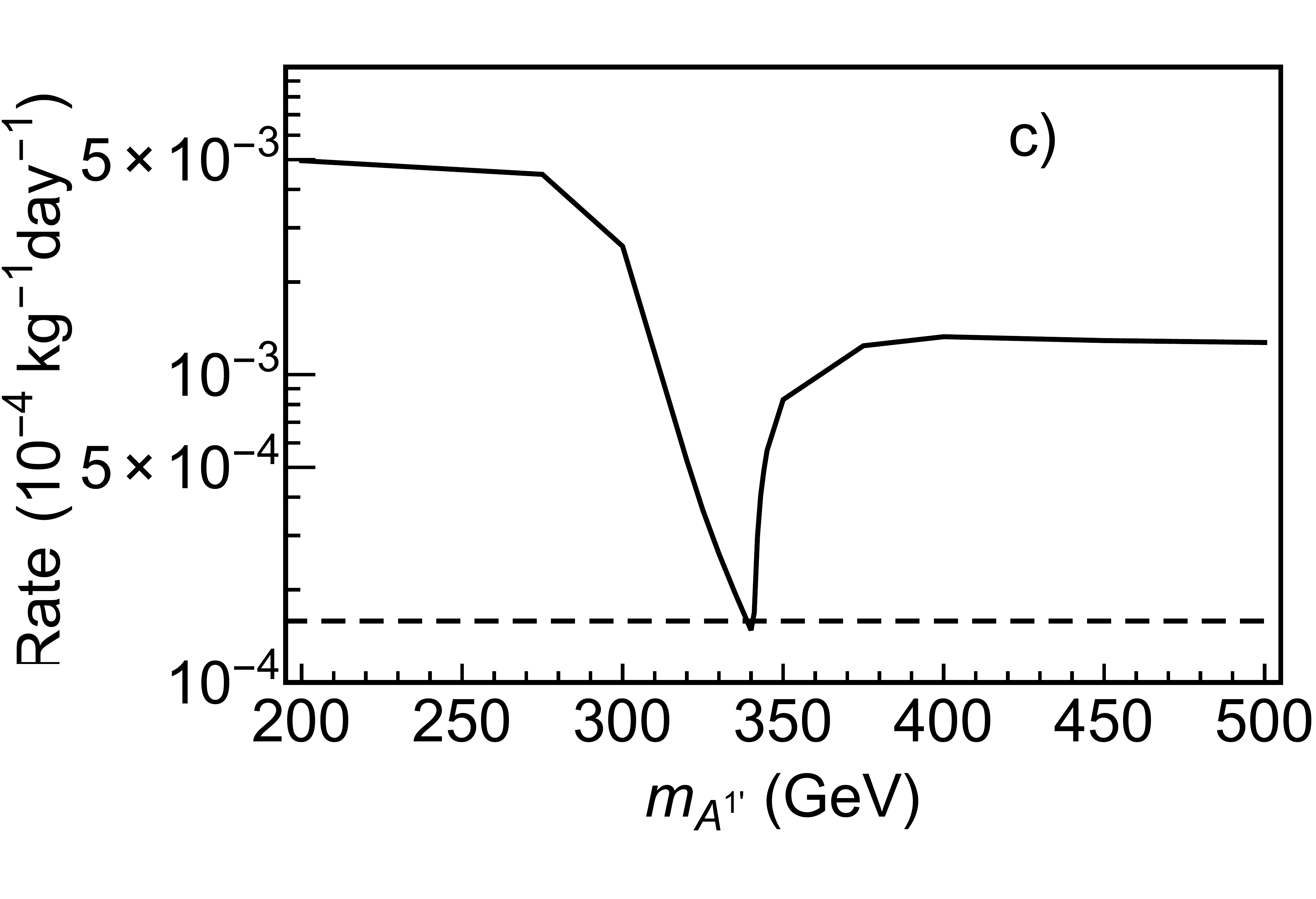}
$ 
$\quad$
$
\includegraphics[scale=0.155,keepaspectratio=true]{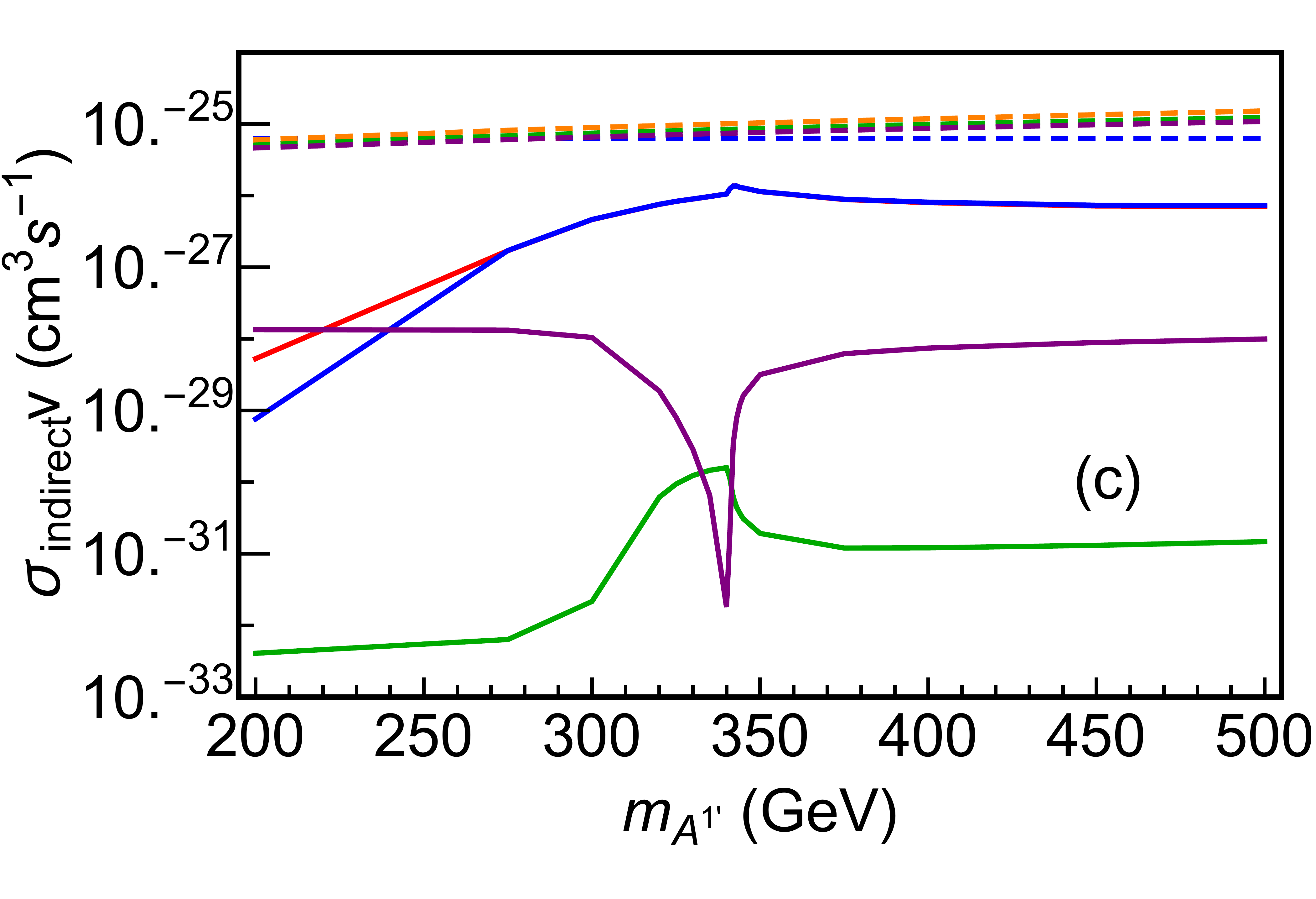}
$ 
\end{center}
\vskip -0.1in
   \caption{ 
   The figures on the left are the
   direct detection event rates as a function of the DM mass 
   and the figures on the right are the indirect detection constraints for the $W^+W^-$ final state.
   In both cases the 
   plots (a), (b), and (c) show the results for benchmark point A, B and C, respectively.
   For direct detection, the allowed points are below the dashed line 
   at $1.58\times 10^{-4}\mbox{ kg}^{-1}\mbox{day}^{-1}$.
   For the indirect detection plots the  line colors represent:
   red for the $\chi\chi\rightarrow W^+W^-$, blue  
   for $\mathcal{H}\mathcal{H}\rightarrow W^+W^-$, 
   green for $A^{1'}A^{1'}\rightarrow W^+W^-$ or $A^{2'}A^{2'}\rightarrow W^+W^-$ which
   have identical results, 
   purple 
   for $A^{3'}A^{3'}\rightarrow W^+W^-$, 
   and orange for $A^{4'}A^{4'}\rightarrow W^+W^-$ or $A^{5'}A^{5'}\rightarrow W^+W^-$ which
   have identical results. The dashed lines are the constraints from Fermi-LAT \cite{Ackermann:2015zua}
   while the solid lines are the predictions 
   of the model. Note that the purple and green dashed line are almost superimposed because they have 
   very similar masses leading to similar constraints. We also note that the solid orange line does not appear
   because it is too small.
}
\label{fig:DDab}
\end{figure}

For DM direct detection, 
we compare our DM direct detection rate limits to the XENON1T measurements \cite{XENON1T:2018}.
Because there are multiple DM species, we need to calculate the expected number of events per species and then sum 
over all species (see Ref.~\cite{KeithBrooks:2012} and \cite{Poulin:2018kap} for details). 
The rate is calculated with the energy range, detection efficiency and exposure 
($1.0\mbox{ t}\times\mbox{yr}$) appropriate for the XENON1T experiment which yields the limit
on the rate of $1.58\times 10^{-4}\mbox{ kg}^{-1}\mbox{day}^{-1}$ at 95\% C.L..  The predicted rate
is dependent on the DM nucleus cross section at zero momentum transfer which in turn is dependent on
the cross section for DM species $i$ and a nucleon.  This cross section is mediated by two scalars, $h_1$ and $h_3$, so
that the cross section for DM species $i$ and a nucleon is given by:
 \begin{align}
\sigma_{in}^0=\frac{f_N^2}{4\pi}\left(\frac{g_{iih_1}c_\theta}{m_{h_1}^2}+\frac{g_{iih_3}s_\theta}{m_{h_3}^2}\right)^2\frac{m_n^2}{(m_n+m_i)^2v^2},
\label{eqn:dd}
\end{align}
where $c_\theta$ and $s_\theta$ are $\cos\theta$ and $\sin\theta$, respectively; 
$f_N\simeq 0.30 \pm 0.03$ 
is the SM Higgs effective coupling to nucleons; 
$m_n$ is the nucleon mass;  
$m_i$ is the mass of the DM species in question; $v$ is the
SM vev; and $g_{iih_1}$ and $g_{iih_3}$ are the couplings between the DM and $h_1$ and $h_3$,
respectively.  The predicted rates along with the XENON1T limit are shown in Fig.~\ref{fig:DDab}. 
The scalar DM, $\chi$ and ${\cal H}$, interact weakly with nucleons because 
the scalar-$h_1$ coupling is small and $h_3$ interacts with nucleons via its SM Higgs component
which is suppressed by the small mixing angle.  As a consequence, when the DM is mainly scalar it 
is weakly constrained by direct detection.

For DM indirect detection, 
we compare our predictions to the DM cross section limits given by the Fermi-LAT collaboration \cite{Ackermann:2015zua} .
Only the $W^+W^-$ final states provided useful constraints \cite{Poulin:2018kap}.  We obtained bounds by 
scaling the predicted cross sections for DM species $i$  by its fractional component of the 
relic abundance:
\begin{align}
\sigma_{\rm scaled}=\left(\frac{\Omega_i}{\Omega_{DM}}\right)^2 
{ { (\sigma_{ii\rightarrow SM+SM} )^2 } \over{ \sigma_{ii}^{tot} }},
\end{align}
where $\sigma_{ii\rightarrow SM+SM}$ is the cross section from two identical DM particles
to one of the SM final states considered by the Fermi collaboration and 
$\sigma_{ii}^{tot} $ is the sum of the cross sections to all possible final states.
The results for each of the benchmark points are shown in Fig.~\ref{fig:DDab}. In all
cases the scalar DM 
has a significantly larger scaled cross section to $W^+W^-$ than
any vector DM because the vector DM annihilates preferentially to other DM particles.  
At present, indirect detection does not impose any constraints although in all three cases, there
is the potential to rule out large regions of parameter space with moderate improvements to the
experimental bounds.

One sees from Fig.\ref{fig:DDab} that for benchmark point B, which is mainly scalar DM, the predicted direct detection
cross sections
are significantly below the XENON1T limits but close enough to the indirect detection limits that they  can
potentially be probed in the foreseeable future.  In contrast,  benchmark C is mainly vector and is almost entirely 
ruled out by direct detection limits.  Benchmark point A is interesting as in some regions it is mainly scalar while 
in other regions mainly vector.  
When the DM is mainly vector it is ruled out by direct detection limits, 
but is far below the indirect detection limits, while when it is mainly 
scalar, the opposite happens.  

\section{Summary}

We reported on a study \cite{Poulin:2018kap} of the hidden SU(3) DM model of Arcadi {\it et al} \cite{Arcadi:2016kmk}
which has multiple vector and scalar DM species.  We used a coupled Boltzmann equation that can
incorporate arbitrary numbers of DM species to calculate the relic abundance \cite{DHT2018}.  We focused on
3 benchmark points with 4 and 7 DM species.  In general, it is important to include all stable particles
in the freeze-out calculations as interactions with species with even small particle densities can
significantly alter the final results. We studied the sensitivity of direct and indirect detection measurements
to the DM and found that vector DM is best constrained by direct detection measurements while scalar DM
is best constrained by indirect detection measurements.  This is a reminder of the complementarity
of the different measurements.  It also points out that with complicated DM sectors, it is possible 
to obtain the correct relic abundance but still not be able to observe it with these measurements. 

\acknowledgments
We thank Keith Dienes and Brooks Thomas for 
helpful conversations and communications.
This work was supported by the Natural Sciences and Engineering Research Council of Canada.

\end{document}